\documentclass[aps,pre,twocolumn,groupedaddress]{revtex4}
\usepackage{graphicx}

\def\giorno{17 July 2006}

\def\b{\beta}

\def\de{\delta}   
\def\eps{\varepsilon}
\def\la{\lambda}
\def\s{\sigma}

\def\om{\omega}

\def\vth{\vartheta}
\def\vphi{\varphi}

\def\pa{\partial}
\def\d{{\rm d}}       

\def\<{\langle}
\def\>{\rangle}

\def\({\left(}
\def\){\right)}
\def\[{\left[}
\def\]{\right]}
\def\=#1{\bar #1}
\def\~#1{\widetilde #1}
\def\.#1{\dot #1}
\def\^#1{\widehat #1}
\def\"#1{\ddot #1}

\def\Y{Yakushevich }

\begin{document}

\title{Solitons in the Yakushevich model of DNA beyond the contact
approximation\thanks{Work supported in part by the Italian MIUR
under the program COFIN2004, as part of the PRIN project {\it
``Mathematical Models for DNA Dynamics ($M^2 \times D^2$)''}.}}

\author{Giuseppe Gaeta}
\affiliation{Dipartimento di Matematica, Universit\`a di Milano,
via Saldini 50, 20133 Milano (Italy); {\tt gaeta@mat.unimi.it}}

\date{\giorno}

\begin{abstract}
The \Y model of DNA torsion dynamics supports soliton solutions,
which are supposed to be of special interest for DNA
transcription. In the discussion of the model, one usually adopts
the approximation $\ell_0 \to 0$, where $\ell_0$ is a parameter
related to the equilibrium distance between bases in a
Watson-Crick pair. Here we analyze the \Y model without $\ell_0
\to 0$. The model still supports soliton solutions indexed by two
winding numbers $(n,m)$; we discuss in detail the fundamental
solitons, corresponding to winding numbers (1,0) and (0,1)
respectively.
\end{abstract}


\maketitle

\section*{Introduction}

Following the pioneering paper and proposal by Englander,
Kallenbach, Heeger, Krumhansl and Litwin \cite{Eng}, a number of
authors considered simple idealized models for the roto/torsional
dynamics of DNA one, with the aim of describing nonlinear
solitonic excitations which -- according to the ideas of Englander
{\it et al.} -- would be related to the transcription bubble which
travels along with RNA-Polymerase in the DNA transcription
process, and would thus play a functional role in this process.

These models are based on modelling the DNA molecule as a double
chain of coupled pendulums; the relevant nonlinear excitations
would then be (topological and dynamical) solitons like those of
the sine-Gordon equation. For a review of the approaches in this
direction, see \cite{YakuBook}; for general properties of DNA and
its functions, see e.g. \cite{CD}.

It should be noted that in a related approach -- but in a
different direction -- the stretching motions of the DNA molecule
(related to DNA denaturation) have also been studied by nonlinear
models, see in particular the Peyrard-Bishop model \cite{PB} and
extensions thereof \cite{BCP,BCPR,PeyNLN}. In this note we will
focus on roto/torsional dynamics, and thus we will not deal with
Peyrard-Bishop like models.

Coming back to rotational dynamics, a very interesting and quite
successful model (also called Y-model) was put forward by prof.
L.V. Yakushevich \cite{YakPLA}. See \cite{YakPhD,YakuBook,YakPRE}
for further results and extensions.

In the \Y model, the DNA molecule is considered homogeneous (i.e.
all nucleotides are considered as identical), and the state of
each nucleotide is described by a single degree of freedom; in
fact, a rotation angle of the base belonging to the nucleotide
around the $C_1$ atom in the sugar-phosphate backbone. Each
nucleotide (or base) is represented as a disk.

Interaction between successive nucleotides on each DNA helix is
via a harmonic potential, while the intrapair interaction (i.e.
interaction between bases in a Watson-Crick pair, and through
these between the corresponding nucleotides) is modelled by a
potential which albeit harmonic in the distance, becomes
anharmonic when described in terms of the relevant degrees of
freedom, i.e. rotation angles. More precisely, with $\ell$ the
distance between relevant points $B_1$ and $B_2$ (see fig.2) of
the disks representing nucleotides in the \Y model, and $\ell_0$
the distance between the points $B_1$ and $B_2$ in the equilibrium
(i.e. the B-DNA) configuration, the intrapair potential is $V_0 =
(1/2) K_p (\ell - \ell_0)^2$, with $K_p$ a dimensional constant.

In the standard \Y model, one considers the approximation $\ell_0
= 0$, which leads to a number of computational simplification; we
call this the {\it contact approximation}, as it corresponds to
having contact between the disks representing nucleotides at same
site on the two chains. However, as pointed out by Gonzalez and
Martin-Landrove \cite{GML}, $\ell_0=0$ is a singular case in that
the description one thus obtains is not structurally stable: as
soon as we consider $\ell_0 \not= 0$, certain qualitative features
of the model dynamics are changed.

In this note we will analyze the \Y model beyond the contact
approximation, i.e. with a nonzero value for the parameter
$\ell_0$; we focus in particular on travelling wave solutions and
soliton-like excitations, as they are the most important objects
in the approach of Englander {\it et al.} \cite{Eng}.

It turns out that in this case soliton solutions are still
present, albeit the simple form obtained with the contact
approximation is replaced by an expression involving elliptic
integrals. The qualitative form of soliton solutions is little
changed with respect to the standard Y-model; as for their width,
it is very moderately increased (see fig.5), still remaining in
the same order of magnitude.

\bigskip\noindent
{\bf Acknowledgements.} I would like to thank M. Cadoni and R. De
Leo for useful discussions, and anonymous referees for
constructive criticisms. This work received support by the Italian
MIUR (Ministero dell'Istruzione, Universit\`a e Ricerca) under the
program COFIN2004, as part of the PRIN project {\it ``Mathematical
Models for DNA Dynamics ($M^2 \times D^2$)''}.

\section{The model}

In the \Y model the DNA double chain is considered to be infinite,
and all bases are considered as having the same physical
characteristics; it is thus an ``ideal'' DNA model according to
Yakushevich's classification \cite{YakPhD,YakuBook}.

Moreover, each nucleotide is considered as a single unit and its
state described in terms of an angular variable. More precisely,
each nucleotide is modelled as a rigid disk which can rotate
around an axis and has a moment of inertia $I$ around this axis,
see fig.1. The rotation of the nucleotide at site $i \in {\bf Z}$
on the chain $a$ ($a=1,2$) is described by an angle
$\vth^{(a)}_i$; we orient all angles in counterclockwise
direction. When we are referring to a specific base pair, we write
simply $\vth_1 , \vth_2$ for $\vth^{(1)}_i,\vth^{(2)}_i$.

\begin{figure}
  \includegraphics[width=150pt]{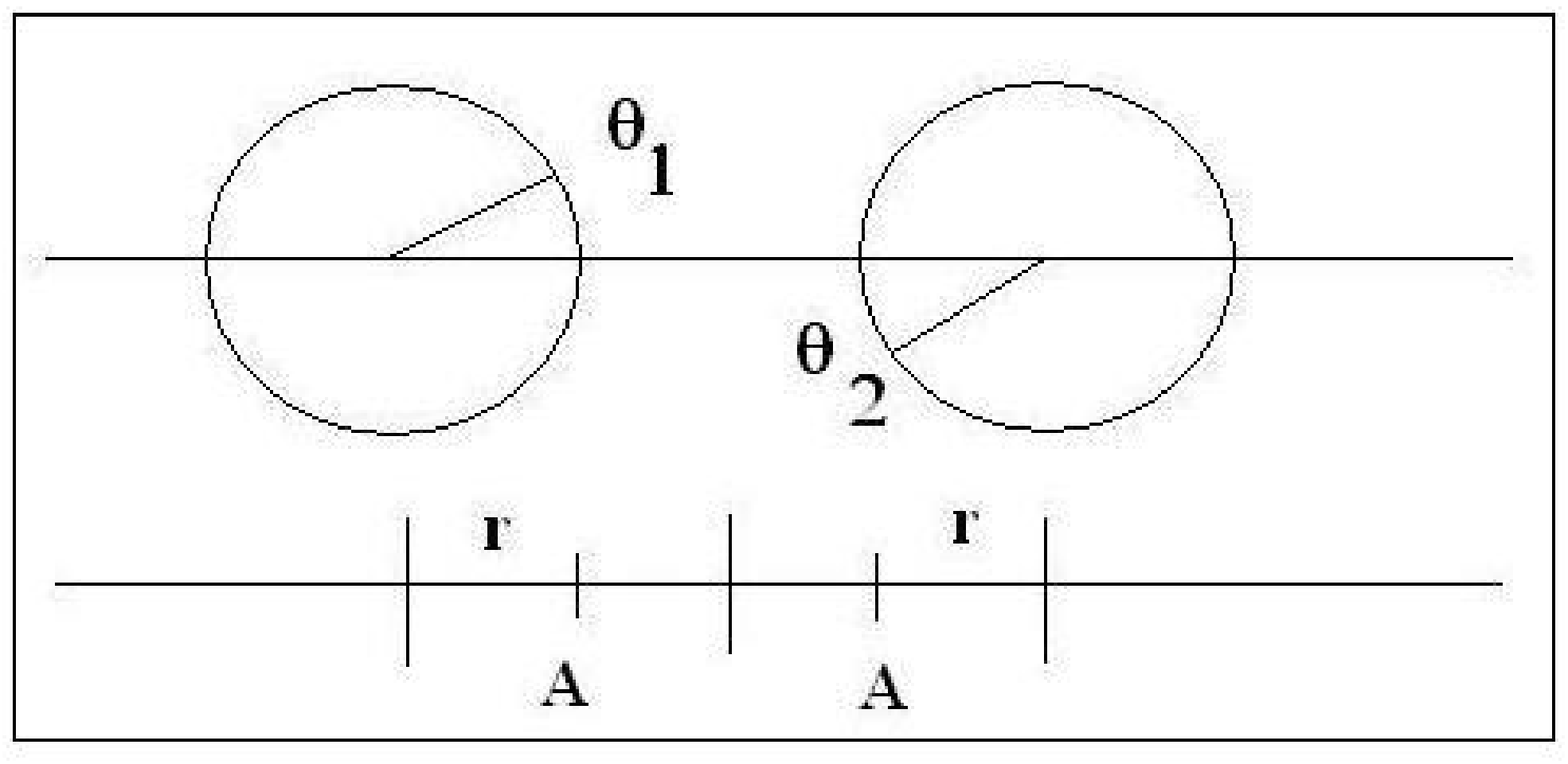}\label{ym1}\hfill
    \includegraphics[width=60pt]{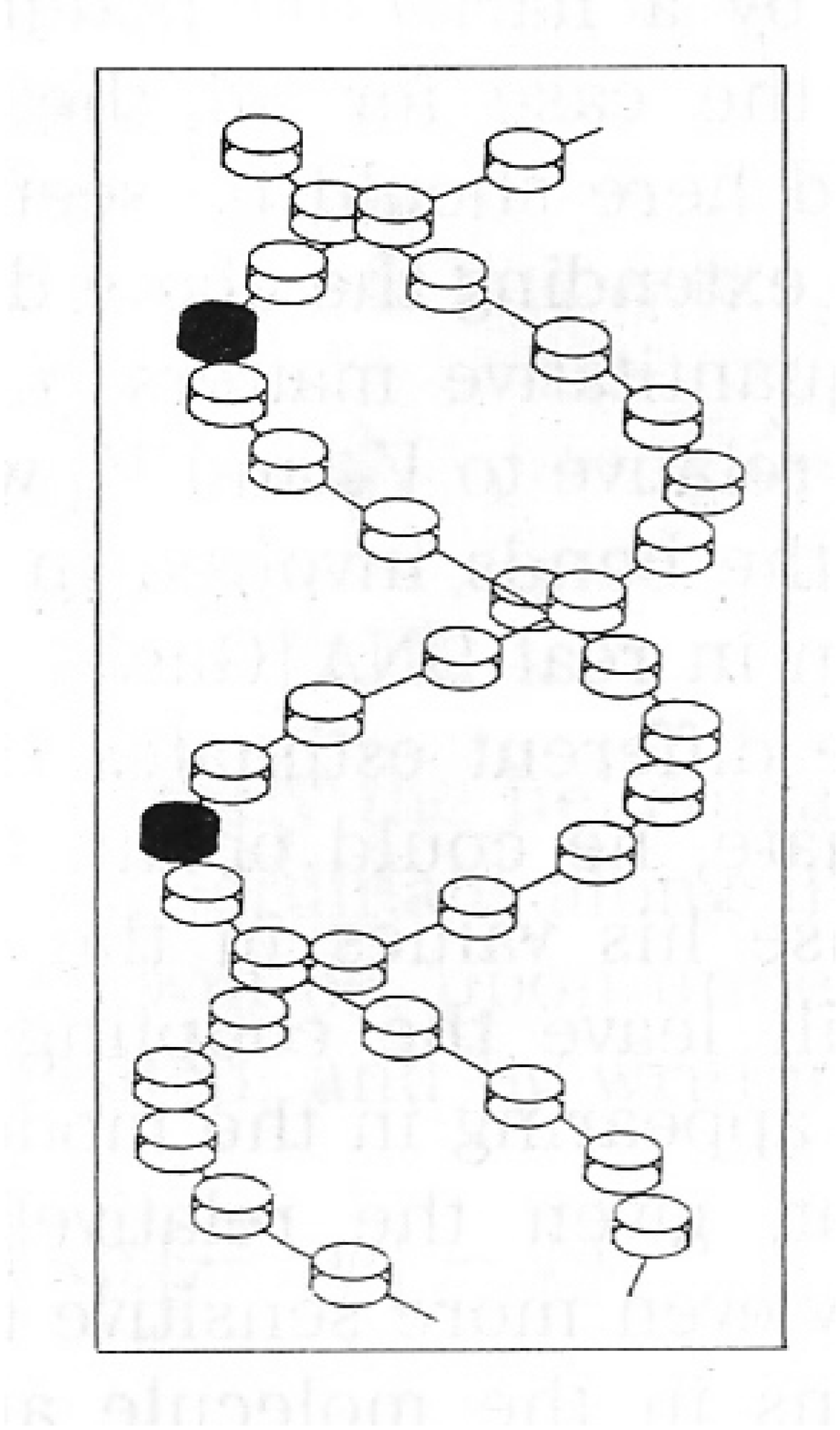}\ \
  \caption{Left: A base pair in the \Y model. The only moving elements in
  nucleotides are the bases; these are represented as identical disks of radius $r$,
  which can rotate around their centers. The disks centers lie at a distance
  $A$ from the double helix axis, and the rotation angles (positive in
  counterclockwise direction) are $\theta_1$, $\theta_2$ respectively.
  The distances $A$ and $r$ are related via $A = r + \ell_0 / 2$, and $\ell_0$
  is the distance between the disks, $\ell_0 = 2 (A - r)$.
  Right: disks represent bases and are located on the double helix;
  two disks as those marked in black interact via the helicoidal terms considered
  in the appendix.}
\end{figure}

\begin{figure}
  \includegraphics[width=200pt]{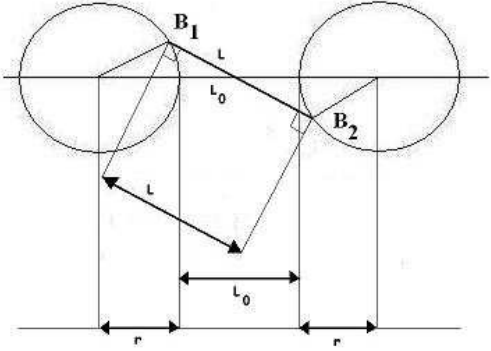}\\
  \caption{Further detail of the base pair modelling in the \Y model, showing in
  particular the points $B_1$ and $B_2$ whose spatial distance is $\ell$. For graphics convenience,
  $\ell$ and $\ell_0$ have been denoted as $L$ and $L_0$ in this plot.}\label{ym2}
\end{figure}


The model is described by a Lagrangian $L = T - U$. The kinetic
energy is
$$  T \ = \ \sum_a \, \sum_i {I \over 2} \, ({\dot \vth}^{(a)}_i )^2 \ . \eqno(1.1) $$

The potential energy $U$ is the sum of two terms; these are the
stacking ($U_s$) term and the pairing one $(U_p$). Thus $ U \, =
\, U_s + U_p$. (There is a third term $U_h$ if we consider the
so-called ``helicoidal'' version of the model
\cite{Dauhel,Gaehel}; introducing $U_h$ leads to qualitative
changes in the small amplitude regime. However, the additional
term $U_h$ is not relevant to fully nonlinear dynamics, and thus
we will not consider it. See the appendix for small amplitude
dynamics, where it matters to introduce this term.)

The first term corresponds to a harmonic potentials (the choice of
harmonic potentials for these term is common to nearly all DNA
models \cite{PeyNLN,YakuBook}) and depend on a dimensional
coupling constant $K_s$. It describes backbone torsion and
stacking interactions between bases at nearby sites on the same
chain; in our notation it is given by
$$ U_s \ = \ {K_s \over 2} \ \sum_a \, \sum_i \, \( \vth^{(a)}_{i+1} -
\vth^{(a)}_{i} \)^2 \ . \eqno(1.2) $$

As for the $U_p$ (pairing) term, it describes the nonlinear
interactions between bases in a pair, i.e. $$ U_p \ = \ \sum_i \
V_0 (\vth^{(1)}_i , \vth^{(2)}_i ) \ . \eqno(1.3) $$ The potential
$V_0$ is not harmonic in terms of the angles $\vth^{(a)}_i$: here
lies the nonlinearity of the dynamics and hence the hearth of the
\Y model. It is assumed that
$$ V_0 \ = \ (1/2) \, K_p (\ell - \ell_0)^2 \ , \eqno(1.4) $$
where $\ell$ is the distance between the atoms which are bridged
by the H bonds -- these are represented in the model by reference
points on the border of the disk representing the bases, and more
precisely by the point which lies nearer to the double helix axis
for $\vth = 0$ -- and $\ell_0$ is this distance in the equilibrium
configuration described by $\vth^{(a)}_i = 0$. We also write $r +
\ell_0 /2 = A$ for the distance between the center of the disks
representing nucleotides and the axis of the double helix.

This harmonic potential becomes anharmonic in terms of the angles
$\vth^{(a)}_i$. Indeed, with simple algebra (and writing $\vth_a$
for $\vth^{(a)}_i$) we have
$$ \begin{array}{rl}
\ell^2 =& (\ell_0^2 + 4 \ell_0 r + 6 r^2 )  -  2 r (\ell_0
+ 2 r)  [ \cos \vth_1 + \cos \vth_2 ] + \\
 & +  2 r^2 \, \cos
(\vth_1 - \vth_2 ) \ . \end{array} \eqno(1.5) $$

In the standard \Y model, one considers the contact approximation
$\ell_0 = 0$, which entails $A=r$; with this, and omitting a
constant term which plays no role, $ U_p  = K_p r^2 \sum_i [ \cos
(\vth^{(1)}_i - \vth^{(2)}_i ) - 2 ( \cos \vth^{(1)}_i + \cos
\vth^{(2)}_i ) ]$. We refer e.g. to \cite{GRPD,YakuBook} for
further detail on the standard \Y model.

In this note we will {\bf not} adopt the contact approximation
$\ell_0 \to 0$, and instead consider the potential energy (1.3)
with $V_0$ given by (1.4), (1.5).

Let us provide an explicit expression for $V_0 (\vth_1,\vth_2)$.
With the notation introduced above, $\ell^2 = (2 A - r \cos \vth_1
- r \cos \vth_2 )^2 + (r \sin \vth_1 + r \sin \vth_2 )^2 $.
Introducing the adimensional parameter $\la = r/A$, this yields
$$ \begin{array}{rl}
V_0 =&  (1/2) K_p (\ell - \ell_0)^2 \\ =& 2 \, K_p \, A^2 \,
\[ (1-\la) - \sqrt{\rho (\vth_1,\vth_2)} \]^2 \ , \end{array} \eqno(1.6) $$
where we have written for short
$$ \begin{array}{rl}
\rho (\vth_1 , \vth_2 ) \ :=& \ \( 1 - \la (\cos \vth_1 + \cos
\vth_2 ) + \right. \\  & \left. + (\la^2/2) [1 + \cos (\vth_1 -
\vth_2 ) ] \) \ .
\end{array} \eqno(1.6') $$ Needless to say, $V_0 (\vth_1 , \vth_2 ) \ge 0$
for all values of $\vth_1 , \vth_2$, the equality being satisfied
only for the equilibrium position $\vth_1 = \vth_2 = 0$.

It should be stressed that this minimum of $V_0$ is non-quadratic,
as remarked by \cite{GML}; indeed, expanding around the
equilibrium $\vth_1 = \vth_2 = 0$, we get
$$ \begin{array}{l}
V_0 (\eps \vth_1 , \eps \vth_2 ) \ = \ (K_p/2) \, \( r [ r (\vth_1
- \vth_2 )^2 + \right. \\ \left. \ - 2 A (\vth_1^2 + \vth_2^2 ) ]
\, / \, [4 (A-r)] \)^2 \ \eps^4 \ + \ O (\eps^6) \ . \end{array}
\eqno(1.7) $$

The equations of motion -- i.e. the Euler-Lagrange equations
arising from the modified Lagrangian -- are
$$ \begin{array}{rl}
I {\ddot \vth}^{(a)}_n =&  K_s  \( \vth^{(a)}_{n+1} - 2
\vth^{(a)}_n + \vth^{(a)}_{n-1} \) + \\ & - \pa V_0
(\vth^{(1)}_n,\vth^{(2)}_n) / \pa \vth^{(a)}_n \ .
\end{array} \eqno(1.8) $$

It is convenient, as in the standard \Y model, to pass to
variables
$$ \psi_n = {\vth^{(1)}_n + \vth^{(2)}_n  \over 2} \ , \
\chi_n = { \vth^{(1)}_n - \vth^{(2)}_n  \over  2} \ ; \eqno(1.9)
$$ these correspond to $\vth^{(1)}_n = \psi_n + \chi_n$ and
$\vth^{(2)}_n = \psi_n - \chi_n$.

In terms of the $\psi , \chi$ variables the eqs. (1.8) read
$$ \begin{array}{rl}
I {\ddot \psi}_n  =&  K_s  \( \psi_{n+1} - 2 \psi_n +
\psi_{n-1} \) \, - \, (1/2) \(\pa V / \pa \psi \) \ , \\
I {\ddot \chi}_n  =&  K_s  \( \chi_{n+1} - 2 \chi_n + \chi_{n-1}
\) \, - \, (1/2) \(\pa V / \pa \chi \) \ .
\end{array} \eqno(1.10) $$
Here we have denoted by $V = V (\psi,\chi)$ the expression of $V_0
(\vth_1,\vth_2)$ in terms of the new variables; that is,
$$ V (\psi , \chi ) \ := \ V_0 (\psi + \chi , \psi-\chi ) \ .
\eqno(1.11) $$

\section{Physical values of the parameters}

Several dimensional parameters appear in our model Lagrangian and
hence in the Euler-Lagrange equations of motion. We will discuss
the model and its predictions for generic values, but specific
values apply to the DNA molecule; for the geometric ones we will
refer for definiteness to B-DNA.

The parameter $A$ represents the distance from the center of
rotation of the disks representing nucleotides to the axis of the
double helix. Taking this center at the $C_1$ atoms (as in the
derivation of parameters in the standard \Y model), we get $2 A =
11.1$\AA for AT pairs and $2 A = 10.8$\AA for GC pairs \cite{Vol}.
We will take the intermediate value $A = 5.48$\AA. (Had we taken
the center of rotation on the phosphodiester chain, $A \simeq
10$\AA would have resulted \cite{YakuBook}).

The choice of the $C_1$ atom as center of rotation for the
nucleotides is not only conformal to discussion of the standard \Y
model \cite{GaeJBP,GRPD,YakuBook}, but also physically sound:
indeed the phosphodiester chain in the backbone is a very flexible
polymer (as also confirmed by the success of Poland-Scheraga type
models based on such flexibility \cite{BLPT,KMP}), while the
complex formed by the sugar ring and the attached base can be seen
in a first approximation as a rigid body \cite{ZC}.

As for $r$ and $\ell_0$, these satisfy $2 A = 2 r + \ell_0$; thus
$r$ is obtained as $r = A - \ell_0 / 2$. Here $\ell_0$ is the
length of the H bond bridging bases in a pair, while $r$ is the
radius of disks representing bases. The physical meaning of $r$
would be the distance from the center of rotation for disks (the
$C_1$ atom in the sugar ring) to the atoms bridged by the
intrapair H bonds.

This distance is quite different from one base to the other, and
also different for different H-bonded atoms in the same base. Thus
we prefer to set the parameter of our idealized model in terms on
the much more uniform value of $\ell_0$, which is $\ell_0 \simeq
2.9 \pm 0.1$\AA  for the different H bonds in Watson-Crick pairs
(see sect.7.2 of \cite{Vol}). We will adopt the value $\ell_0 =
2.9$\AA. With these choices, $r \simeq 4.0$\AA, and hence $\la = r
/ A = [1 - \ell_0/(2A)] = 0.74 \simeq 3/4$.

Finally, $\de$ represents the interbase distance along the double
helix axis; in B-DNA this is $\de = 3.4$\AA.

After discussing the geometrical parameters, let us now come to
the dynamical ones. First of all, we have the moment of inertia
$I$; this is rather different for different bases (detailed values
are given e.g. in \cite{ZC}). Taking an average over different
values, we will adopt as in \cite{GaeJBP} the value $I= 3 \cdot
10^{-37} \, {\rm cm}^2 \, {\rm g}$. As for the coupling constants
$K_s$ and $K_p r^2$ (and $K_h$ considered in the appendix), we
will also adopt the values given in \cite{GaeJBP}, i.e. $K_s =
0.13$eV, $K_p r^2=0.025$eV, $K_h = 0.009$eV.

In our discussion of soliton solutions and their width, we used
the parameters $\mu$ and $B$; the first of these depends on the
speed of the soliton, which is a free parameter (provide it is
smaller than the limiting value $\sqrt{K_s/I} \de$, see sect.3) in
the \Y model. For $v=0$, we get $\mu = - K_s \de^2$. The parameter
$B^{-1} = b \de$ is defined in terms of $\mu$ by (5.8) below. With
our choices for the dimensional parameters, and $\mu$ as above, we
get $ b \simeq (3/8) 2.28 \simeq 0.86$, and hence $B^{-1} \simeq
0.86 \de$, $B \simeq 0.34 \AA^{-1}$.

\section{Continuum approximation}

In discussing (1.10), it is convenient to promote the arrays $\{
\psi_n (t) \}$ and $\{ \chi_n (t) \}$ ($n \in {\bf Z}$) to fields
$\psi (x,t)$ and $\chi (x,t)$ (no confusion should be possible
between old dependent variables and fields), the correspondence
being given by
$$ \psi_n (t) \ \simeq \ \psi (n \de , t ) \ , \ \ \chi_n (t) \
\simeq \ \chi (n \de , t) \ . \eqno(3.1) $$ Here $\de$ is the
spacing between successive sites of each chain; in B-DNA we have
$\de = 3.4$\AA. (It is of course also possible to pass to
adimensional units in the spatial variable, i.e. set $\xi = x /
\de $, and consider $\psi (\xi,t)$, $\chi (\xi,t)$ so that e.g.
$\psi_n (t) = \psi (\xi,t)$ for $\xi = n$.)

With this, (1.10) reads
$$ \begin{array}{rl}
I \psi_{tt} (x,t)  =&  K_s  \( \psi (x+ \de,t) - 2 \psi (x,t)
+ \psi (x - \de , t)  \)  + \\
 & \ - \ (1/2) \(\pa V / \pa \psi \) (x,t) \ , \\
I \chi_{tt} (x,t)  =&  K_s  \( \chi (x+ \de,t) - 2 \chi (x,t) +
\chi (x - \de , t)  \)  + \\ & \ - \ (1/2) \(\pa V / \pa \chi \)
(x,t) \ . \end{array} \eqno(3.2) $$

If now we assume that $\psi (x,t)$ and $\chi (x,t)$ vary slowly in
space compared with the length scale set by lattice spacing, we
can write
$$ \begin{array}{l}
\psi (x \pm \de , t) = \psi (x,t) \pm \de \psi_x (x,t) + (\de^2 /
2) \psi_{xx} (x,t) \ , \\
\chi (x \pm \de , t) = \chi (x,t) \pm \de \chi_x (x,t) + (\de^2 /
2) \chi_{xx} (x,t) \ . \end{array} \eqno(3.3) $$

Inserting these into (3.2), we obtain the equations of motion for
the \Y model in the continuum approximation:
$$ \begin{array}{rl}
I \psi_{tt} (x,t)  =&  K_s  \de^2  \psi_{xx} (x,t)
 -  (1/2) \(\pa V  / \pa \psi \) (x,t) \ , \\
I \chi_{tt} (x,t)  =&  K_s  \de^2 \chi_{xx} (x,t)  -  (1/2) \(\pa
V  / \pa \chi \) (x,t) \ . \end{array} \eqno(3.4) $$ We omit from
now on to specify at which point functions should be evaluated, as
we got a local formulation. We will also consider, where
appropriate, $\vth_1$ and $\vth_2$ as fields.

The PDEs (3.4) should be supplemented with a side condition
specifying the function space to which acceptable solutions
belong. The physically natural condition is that of {\it finite
energy}; that is, we should require that the integral
$$ \int_{-\infty}^{+\infty} \[ {1 \over 2} \( I (\psi_t^2 + \chi_t^2 ) + K_s
(\psi_x^2 + \chi_x^2) \) + {1 \over 2} V (\psi , \chi ) \] \d x
\eqno(3.5) $$ is finite; if this condition is satisfied at $t=0$,
it will be so for any $t$.

It should be mentioned that the \Y equations (3.4) correspond, in
the $\ell_0 = 0$ approximation, to classical ones when only one
field is nonzero: indeed for $\chi = 0$ they reduce to the
sine-Gordon equation, while for $\psi = 0$ one gets the so-called
"double sine-Gordon" equation, which appears in many physical
contexts \cite{BCG,CDeg}.

\section{Travelling wave solutions}

Next we focus on travelling wave solution for (3.4). That is, we
restrict (3.4) to a space of functions
$$ \psi (x,t) \ = \ \vphi (x - v t) \ , \ \chi (x,t) \ = \ \eta (x -
v t) \eqno(4.1) $$ (we will further restrict this in the
following, in order to take into account the finite energy
condition). We will also write simply $z := x - v t$, and
introduce the parameter
$$ \mu \ := \ I \, v^2 \ - \ K_s \, \de^2 \ . \eqno(4.2) $$
Note that $\mu $ could be negative; this will actually be the
interesting case.

With (4.1) and (4.2), the (3.4) reduce to two ODEs, i.e.
$$ \begin{array}{rl}
\vphi'' \ =& \ - \, [1/(2 \mu)] \, \(\pa V (\vphi,\eta) / \pa \vphi \) \ , \\
\eta'' \ =& \ - \, [1/(2 \mu)] \, \(\pa V (\vphi,\eta) / \pa \eta
\) \ . \end{array} \eqno(4.3) $$ These describe the motion (in the
``time'' $z$) of a point particle of unit mass in the effective
potential
$$ W (\vphi , \eta ) \ := \ (2 \mu)^{-1} \ V (\vphi , \eta )
\ . \eqno(4.4) $$ The conservation of energy reads then
$$ {1 \over 2} \( (\vphi')^2 + (\eta')^2 \) \ + \ W(\vphi ,
\eta ) \ = \ E \ . \eqno(4.5) $$

The finite energy condition (3.5) implies, in terms of $\vphi
(z)$, $\eta (z)$, that
$$ \lim_{z \to \pm \infty} \vphi' (z) \ = \ \lim_{z \to \pm \infty} \eta' (z) \ =
\ 0 \ ; \eqno(4.6') $$ moreover, the functions $\vphi$ and $\eta$
themselves should go to a point of minimum for the potential $V$.

If $\mu >0$, the minima of $V$ are the same as the minima of $W$;
but if $\mu < 0$, then minima of $V$ are the same as the maxima of
$W$. As it is impossible that nontrivial motions reach
asymptotically in time (that is, in $z$) a minimum of the
effective potential -- while they can reach asymptotically a
maximum if they have exactly the correct energy -- in order to
have travelling wave solutions satisfying (4.6') and going to
minima of $V$ for $z \to \pm \infty$ we need
$$ \mu \ < \ 0 \ ; \eqno(4.7)
$$ we assume this from now on. Note that $\mu < 0$ implies there
is a maximum speed for travelling waves:
$$ |v| \ < \ \sqrt{K_s/I} \, \de \ . \eqno(4.8) $$

The minima of $V$, i.e. the maxima of $W$, are for
$(\vth_1,\vth_2) = (2 q_1 \pi , 2 q_2 \pi)$; writing $n = (q_1 +
q_2)/2$, $m = (q_1 - q_2)/2$, these correspond to $ (\vphi,\eta) =
(2 n \pi , 2 m \pi)$. Thus the finite energy condition requires
(with obvious notation)
$$ \lim_{z \to \pm \infty} \vphi (z) \ = \ 2 n_\pm \pi \ , \ \lim_{z
\to \pm \infty} \eta (z) \ = \ 2 m_\pm \pi \ . \eqno(4.6'') $$ We
can and will always take $n_- = m_- = 0$ with no loss of
generality; we will hence write simply $n,m$ for $n_+,m_+$. These
satisfy $n - m \in {\bf Z}$ (in addition to $2 n \in {\bf Z}$, $2
m \in {\bf Z}$).

In terms of the dynamical system describing the evolution in
``time'' $z$ in the potential $W$, the solutions satisfying (4.6)
represent heteroclinic solutions connecting the point $(0,0)$ at
$z = - \infty$ with the point $(2 \pi n , 2 \pi m )$ at $z = +
\infty$. It is thus no surprise that the analytic determination of
such solutions is in general impossible.

On the other hand, the solutions with indices $(1,0)$ and $(0,1)$
can be determined explicitly, as shown in the next section.

\section{Special solutions}

The solution with indices $(1,0)$ and $(0,1)$ are special in that
they require that only one of the two fields varies. That is, the
$(1,0)$ solution will have $\eta (z) \equiv 0$; and the $(0,1)$
solution will have $\vphi (z) \equiv 0$. Thus, they correspond to
one-dimensional motions in the effective potential $W
(\vphi,\eta)$, and as such they can be exactly integrated.

Note that these are immediately taken back to a description in
terms of the original angles $\vth_i$: indeed, by (1.9), for the
(1,0) solution we will have $\vth_1 = \vth_2 = \vphi$, while for
the (0,1) solution it results $\vth_1 = - \vth_2 = \eta$.

\subsection{The (1,0) solution}

Setting $\eta = 0$, the effective potential reduces to
$$ P (\vphi ) \ := \ W (\vphi , 0) \ = \ (2 \mu)^{-1} \, V(\vphi,0) \ ; \eqno(5.1) $$
note that $\mu < 0$ implies $P (\vphi) \le 0$ for all $\vphi$. The
first of (4.3) reduces to $ \vphi'' = - [\pa P (\vphi ) / \pa
\vphi ]$, and the conservation of energy (4.5) reads
$$ (1/2) (\vphi' )^2 \ + \ P(\vphi) \ = \ E \ . \eqno(5.2) $$
By construction the solutions satisfying the side conditions (4.6)
correspond to $E = P(0)$, as $\phi (-\infty ) = 0$; moreover,
again by construction,
$$ P(0) \ = \ W(0,0) \ = \ (2 \mu)^{-1} \, V(0,0) \ = \ 0 \ . \eqno(5.3) $$
Thus, the separable equation (5.2) yields
$$ {\d \vphi \over \d z} \ = \ \sqrt{- 2 P(\vphi)} \ . \eqno(5.4) $$
(In the (1,0) solution, $\vphi' >0$; i.e. we have the positive
determination of the root at all $z$.)

The expression for $P$ is readily obtained once we note that $\eta
= 0 $ means $\vth_1 = \vth_2$, see (1.9). With this, the
expression (1.6) for $V_0 (\vth_1,\vth_2)$ gets simplified. We
write, for ease of notation, $$ \begin{array}{l} h (\vphi) :=
\sqrt{1 - 2 \la \cos \vphi + \la^2 } \ , \\ \b (\vphi ) \ := \ \(
(\la -1) \, + \, h(\vphi) \) \ ; \end{array} \eqno(5.5) $$ note
that $h(\vphi)>0$ and $\b (\vphi) \ge 0$ for all $\vphi$ (equality
applying only at $\vphi = 2 k \pi$) for $0<\la<1$. It results with
standard algebra that
$$ V_0 (\vth , \vth ) \ = \ 2 \, K_p A^2 \ \b^2 (\vth) \ . \eqno(5.6) $$
It follows immediately from (5.6) and (4.4), (5.4) that $\vphi$ is
obtained by integrating
$$ {\d \vphi \over \b (\vphi ) } \ = \ B \ \d z \eqno(5.7) $$
with $B$ a constant, $$ B := \ A \sqrt{- 2 K_p / \mu} \ = \ A \,
\sqrt{2 K_p /|\mu|} \ . \eqno(5.8) $$ That is, $\b (\vphi )$
represents the rate of variation with $z$ of the soliton solution
in adimensional units, set by (5.8).

The integral $f(\vphi)$ of the left hand side of (5.7) is given
explicitly by
$$ \begin{array}{rl}
f(\vphi) \ =& \ c_0 \ - \ \[2 \la \]^{-1} \, \times \, \[ (1 - \la
) \, E (\vphi/2,\s) \, + \right. \\
 & \left. \ - \, [(1+\la)^2
/ (1-\la)] \, F (\vphi/2,\s) \, + \right. \\
 & \left. + \, [1 - \la + h(\vphi)] \,
\cot (\vphi/2) \] \ .
\end{array} \eqno(5.9) $$

Here $F$ and $E$ are the elliptic integrals of first and second
kind respectively, defined as $F(x,\s) = \int_0^x (1 - \s \sin^2
\theta)^{-1/2} \d \theta$ and $E(x,\s) = \int_0^x (1 - \s \sin^2
\theta)^{1/2} \d \theta$. The complete elliptic integrals are
${\mathcal K}(\s) = F(\pi/2,\s)$ and ${\mathcal E}(\s) =
E(\pi/2,\s)$. In (5.9) we have moreover $\s = - 4 \la /(1-\la)^2$,
and $c_0$ is the integration constant. The latter can and will be
chosen as
$$ c_0 \ = \ {(1-\la)^2 \, {\mathcal E} (\s ) \ - \ (1+\la)^2 \,
{\mathcal K} (\s ) \over 2 \, \la \, (1-\la )} \ , \eqno(5.10) $$
with ${\mathcal K}$ and ${\mathcal E}$ the complete elliptic
integrals of first and second kind respectively; in this way
$f(\vphi)$ is antisymmetric with respect to $\vphi = 0$. The
function $f(\vphi)$ is singular at $\vphi = 2 n \pi$, as seen in
fig.2.

\begin{figure}
    \includegraphics[width=200pt]{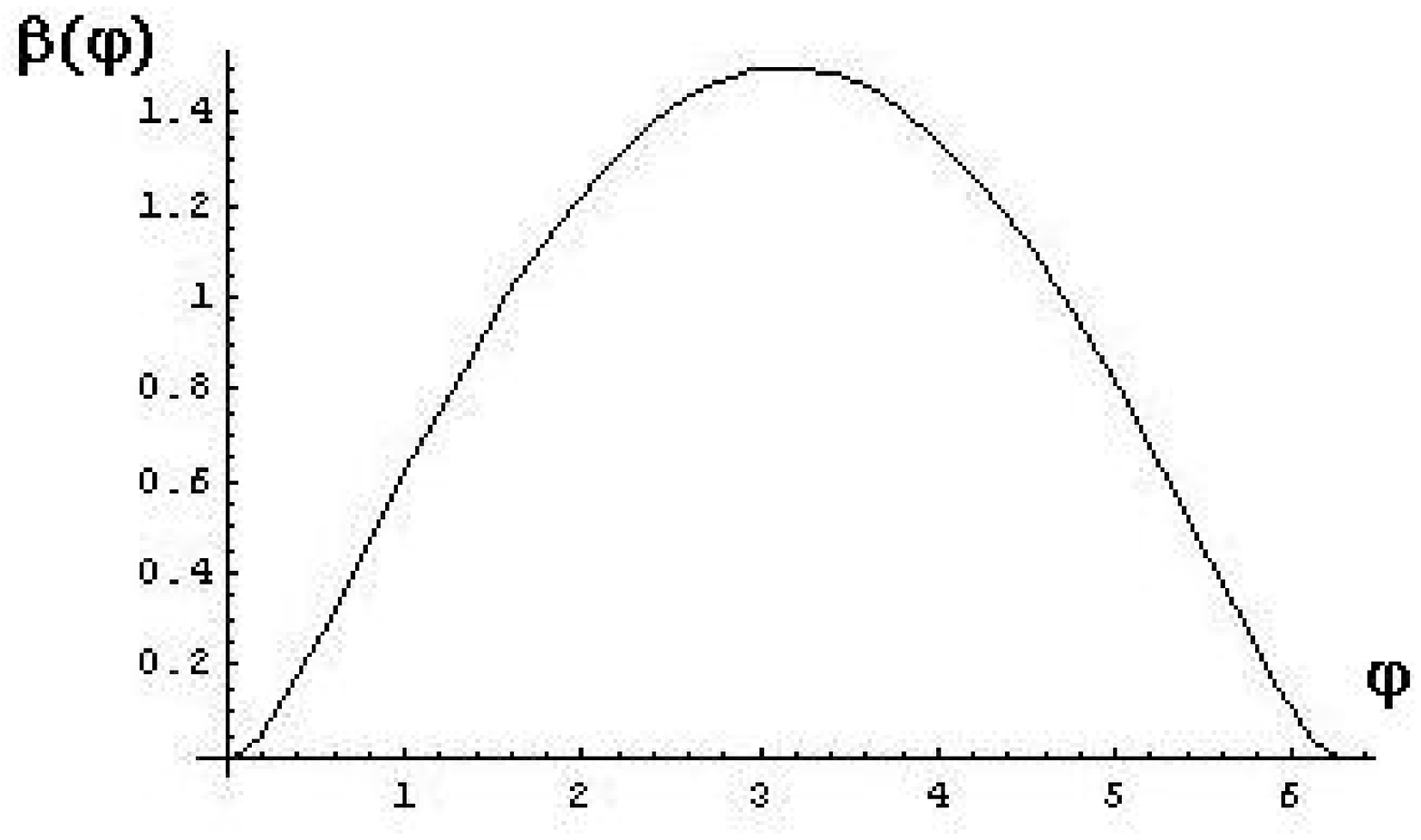}\\
    \includegraphics[width=200pt]{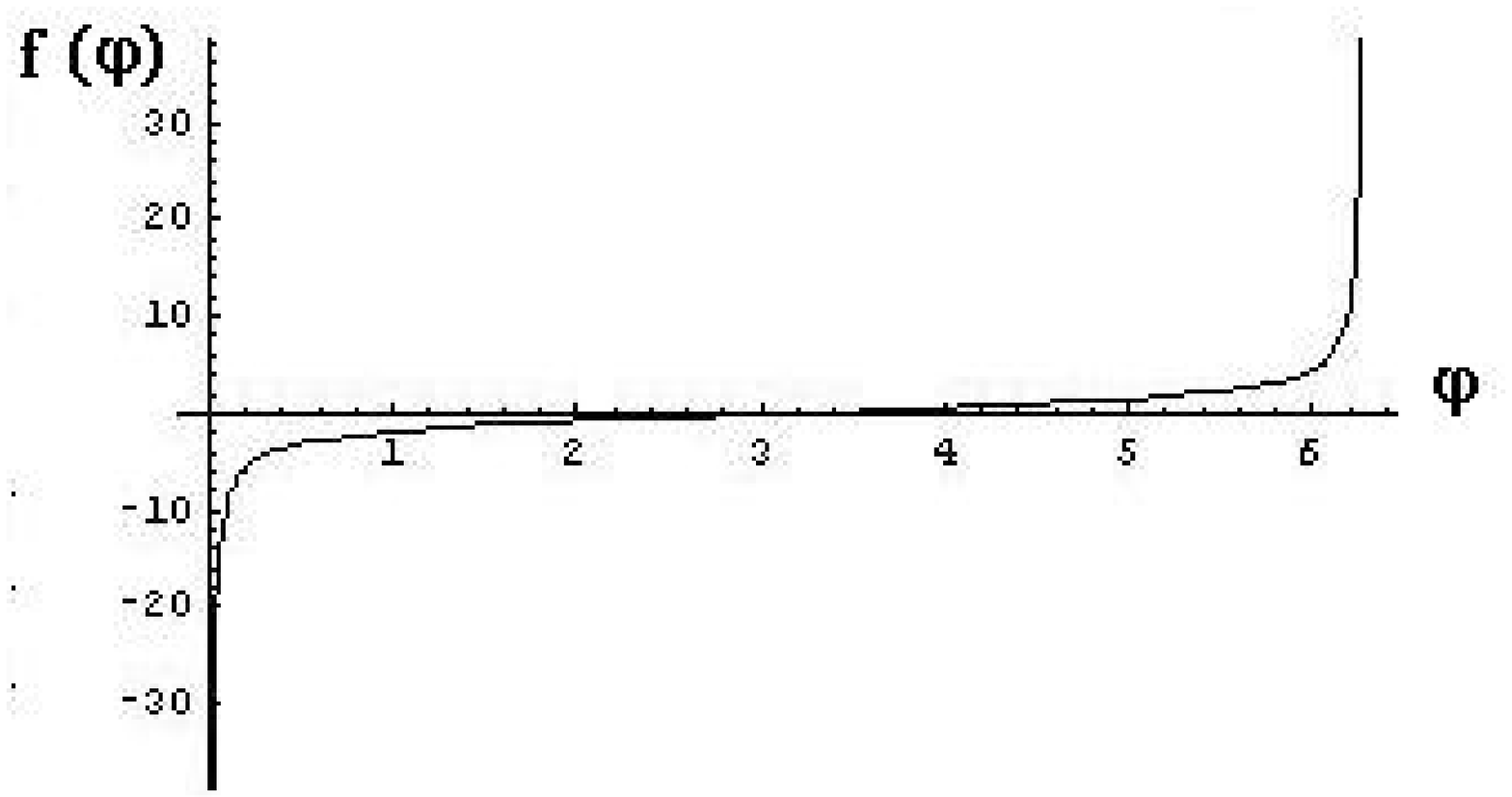}\\
  \caption{Upper plot: the function $\b (\vphi)$, representing the
  "speed" $\d \vphi / \d z$ (in terms of the effective dynamics description provided by (5.4))
  of the (1,0) soliton in adimensional units (see text and (5.7)), for $\la = 3/4$.
  Lower plot: the function $f(\vphi) = \int [1/\b (\vphi)] \d
  \vphi$,to be inverted in order to get the (1,0) soliton solution
  (see 5.11)), again for $\la = 3/4$; here $c_0$ has been chosen according to (5.10),
  so that $f (\pi) = 0$.}
\end{figure}

Needless to say, integrating also the right hand side of (5.7), we
get $ f(\vphi ) = B (z - z_0 ) $. With $z_0 = 0$ (this integration
constant can be absorbed in $c_0$), this yields finally for the
(1,0) solution (shown graphically in fig.3)
$$ \vphi \ = \ f^{-1} (B z) \ . \eqno(5.11) $$

\begin{figure}
  \includegraphics[width=200pt]{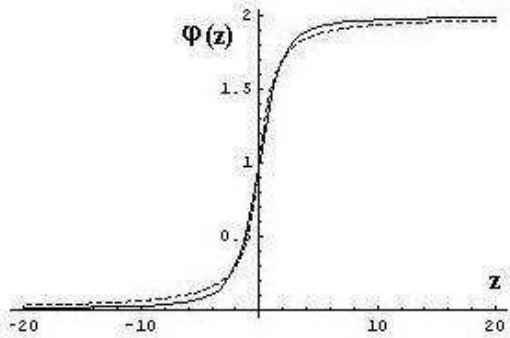}\\
  \caption{The (1,0) soliton for the \Y model without the contact approximation,
  see (5.11), with $\la=3/4$ (solid lines); and for the \Y model
  with the contact approximation (dotted lines).
  Here $\vphi$ is in units of $\pi$, and $z$ in units of the distance $\de$ between
  successive base pairs.}
\end{figure}

\subsection{The (0,1) solution}

For the (0,1) solution we can set $\vphi = 0$, which means $\vth_2
= - \vth_1$. With this, and writing again $r = \la A$,
$$ (\ell - \ell_0)^2 \ = \ 4 \ A^2 \ \la^2 \ \(1 - \cos \eta \)^2
\ . \eqno(5.12) $$ The effective potential is hence given by
$$ Q (\eta ) = W(0,\eta) = { V(0,\eta) \over 2 \mu}  =
{(A \la)^2 K_p \over \mu} (1 - \cos \eta )^2 \ . \eqno(5.13)
$$
Note that again $\mu < 0$ implies $Q(\eta) \le 0$ for all $\eta$,
and that $Q(0) = 0$. Conservation of energy provides in this case
$\d \eta / \d z = \sqrt{- 2 Q (\eta)}$; hence we have
$$ {\d \eta \over \d z} \ = \ \la \, B \, (1 - \cos \eta ) \eqno(5.14) $$
with $B$ as above. Equation (5.14) is immediately integrated,
providing
$$ {\rm ctg} (\eta / 2) \ = \ - \, \la \, B \, (z-z_0) \ . \eqno(5.15) $$
We can choose $z_0= 0$, so that $\eta (\pi) = 0$, and $\eta$ is
antisymmetric with respect to $z=0$. In conclusion, the (0,1)
solution is given by
$$ \begin{array}{rl}
\eta  =&  - 2 \, {\rm arcctg} (\la B z) \ = \\
 =&  2 \, \[ \pi -
\arccos\({\la B z \over \sqrt{1 + \la^2 B^2 z^2} }\)
\] \ ; \end{array} \eqno(5.16) $$ this is shown in fig.4.
Note that some care should be taken in using
appropriate determination of the ${\rm arcctg}$ and $\arccos$
functions so that $\eta$ is continuous at $z=0$.

\begin{figure}
  \includegraphics[width=200pt]{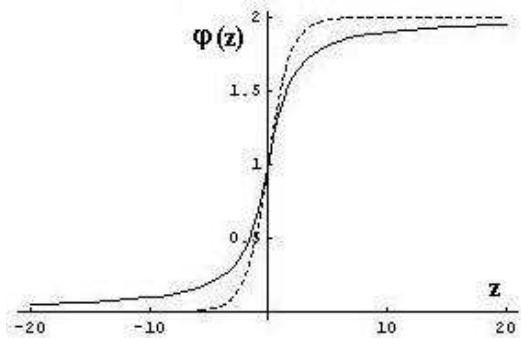}\\
  \caption{The (0,1) soliton
  for the \Y model without the contact approximation,
  see (5.5), with $\la=3/4$ (solid lines); and for the \Y model
  with the contact approximation (dotted lines); units as in fig.3.}
\end{figure}

\subsection{Comparison with standard Y solitons}

The standard \Y model predictions are recovered for $r = A$, i.e.
for $\la = 1$. The solitons shape is evidently very similar to
those of standard \Y solitons also for $\la \not=1$, see figs.3
and 4.

In order to compare more precisely the results obtained within and
without the contact approximation, we recall that with the contact
approximation the (1,0) and (0,1) \Y solitons are given
respectively by \cite{GRPD,YakuBook}
$$ \begin{array}{lr}
\vphi (z) \ = \ 4 \, \arctan \, [e^{B z} ] &, \eta = 0 \ ; \\
\eta (z) \ = \ 2 \, \arccos \, \[ - B z / \sqrt{1 + B^2 z^2 } \]
&, \vphi = 0 \ . \end{array} \eqno(5.17) $$

As for the soliton width, which represents the size of the
transcription bubbles in the Englander {\it et al.} theory, this
can be determined exactly via (5.9) and (5.15) once we decide how
this should be measured. That is, the width will correspond to
$z_+ - z_-$ where $z_\pm$ are the points at which the angles
$\vphi$ or $\eta$ differ by their asymptotic value ($0$ or $2
\pi$) by less than a given amount $\Delta$. This is shown in fig.5
for small values of $\Delta$.

It should be stressed that the soliton widths vary quite slowly
with $\la$; thus, the predictions of the model will not
sensitively depend on the precise value of $\la$.

More precisely, these are given by the parameter $B^{-1}$, which
-- see (5.8) -- is written as
$$ B^{-1} \ = \ (\la /2) \ \sqrt{|\mu|/(K_p r^2)} \ ; \eqno(5.18) $$
again we stress that the curves plotted in fig.5 have a not so
steep slope, which shows that the predictions of the model do not
depend too much on the precise value of $B$.

In the limit $v \to 0$ we have $|\mu| = K_s \de^2$ and hence
$$ B^{-1} \ = \ b \ \de \ \ {\rm with} \ \
b \ = \ {\la \over 2} \, \sqrt{K_s \over K_p r^2} \ . \eqno(5.19)
$$

As clear from fig.5, dropping the contact approximation will cause
a widening of the \Y solitons; this goes in the right direction
since the standard \Y model produces the right order of magnitude
for the soliton width but with a too small exact numerical value
\cite{GaeJBP}.

Indeed, it is experimentally known that the transcription bubbles
have a width of about 15--20 base pairs; this is the size of the
region in which the base pairs are open and the base sequence can
be accessed by the RNA Polymerase. It is not easy to assess what
is precisely the angle at which base pairs should be considered as
open, so we have plotted different possibilities in fig.5.

\begin{figure}
  \includegraphics[width=200pt]{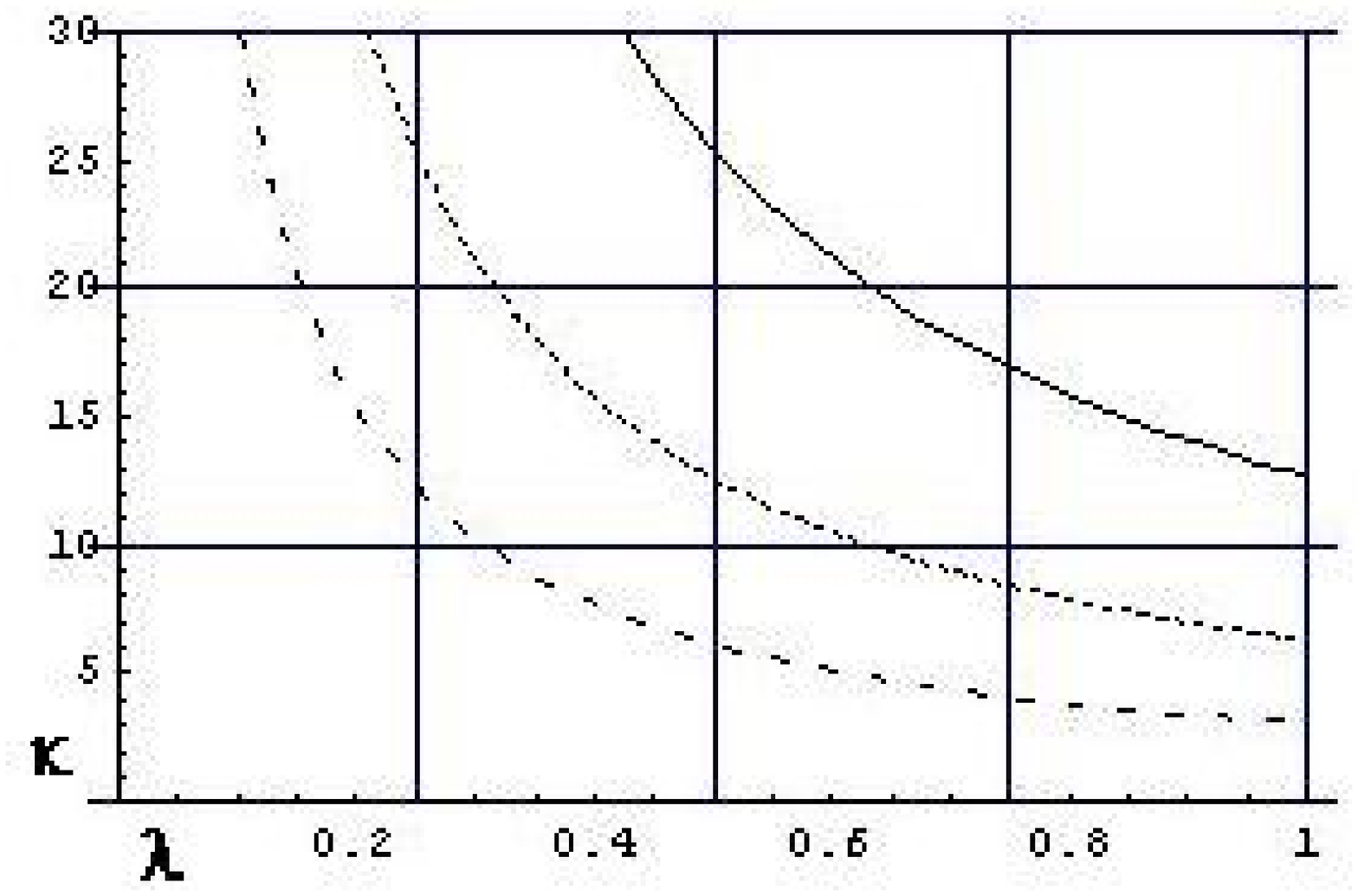}\\
  \includegraphics[width=200pt]{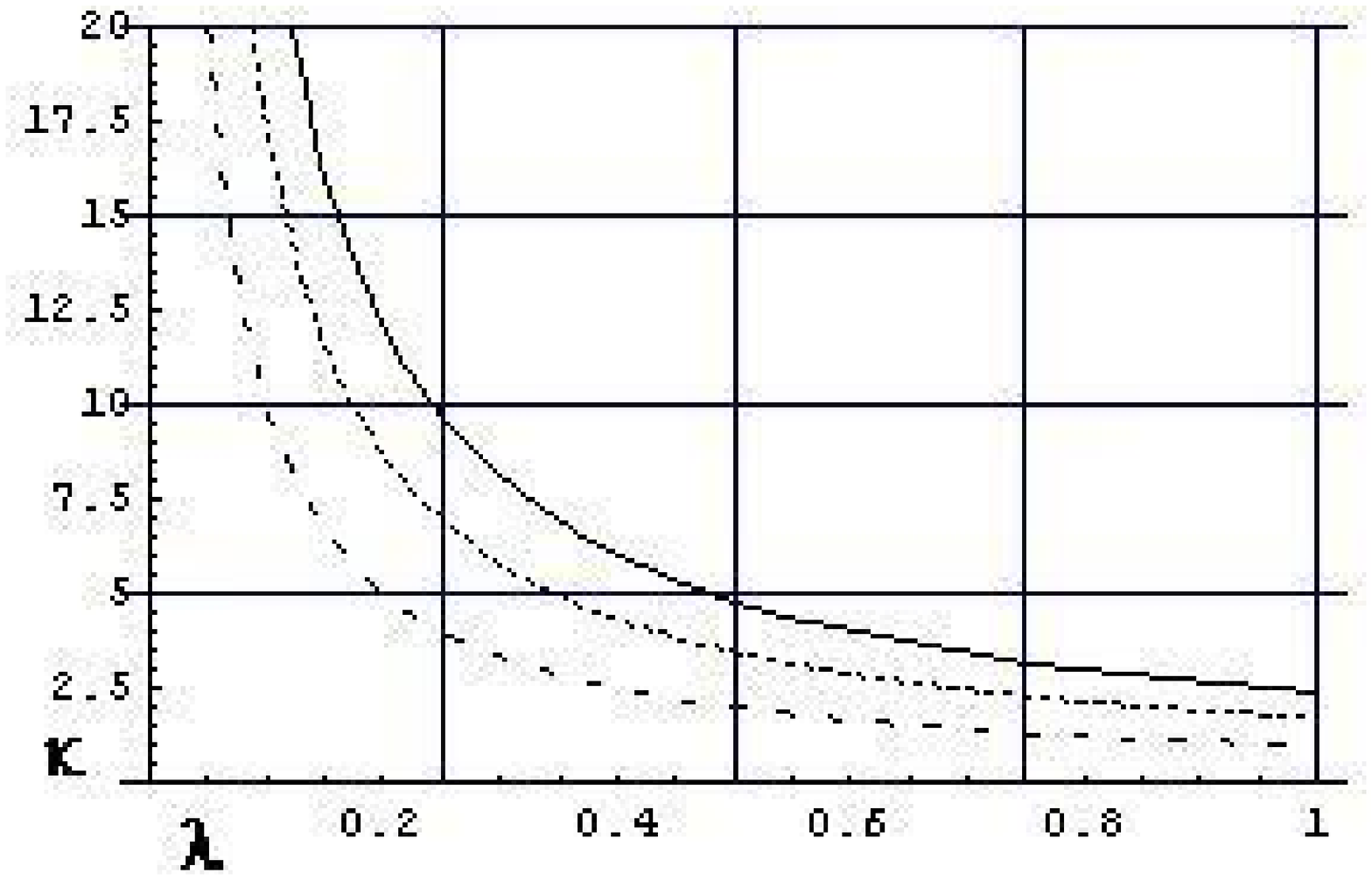}\\
  \caption{Half-width for the (1,0) soliton (upper) and the (0,1) soliton (lower)
  for the \Y model without the contact approximation. The curves
  show the half-width $\kappa$ (in units of $B^{-1}=b \de$) of the solitons as
  functions of $\la$, with different conventions for the measure
  of the half-width itself, see text: $\Delta=\pi/20$ (continuous curves),
  $\Delta = \pi/10$  (dotted curves) and $\Delta = \pi/5$ (dashed curves).}
\end{figure}

\section{Conclusions and outlook}

We have considered the \Y model beyond the contact approximation
$\ell_0 \to 0$, focusing on the solitonic excitations which --
according to Englander {\it et al.} -- are supposed to play a
functional role in the transcription process.

We have shown that in this case soliton solutions are still
present, and can be described exactly in analytical terms; the
simple form obtained within the contact approximation is replaced
by a more complex expression, involving elliptic integrals.
However, the qualitative form of soliton solutions is little
changed, and their width is only very moderately increased.

This shows that the \Y model is actually quite robust against
changes in $\ell_0$ and dropping of the contact approximation;
this not really for what concerns its mathematical aspects, but
rather for what concerns its {\it physical} features and
predictions, in particular in the fully nonlinear regime.

Thus, the first outcome of our work is that one is physically
quite justified in considering the simplifying contact
approximation in the \Y model, albeit the analysis can be
performed with the same completeness without that approximation.

Let us mention that other work in progress \cite{CDG} show that by
considering a more detailed description -- in various ways -- of
the DNA molecule, one obtains indeed new features with respect to
simple idealized models as the \Y one.

In this respect, the present work suggests that in analyzing these
more detailed -- and hence more difficult to study -- models one
can in the first instance adopt the same kind of approximation
adopted by \Y in her original study, and focus instead on other
features of the model. This suggestion will be taken up in
forthcoming work \cite{CDG}.


\section*{Appendix. Dispersion relations}

In this note we are mainly interested in solitonic excitations,
hence fully nonlinear dynamics. However, the study of small
amplitude excitations has some interest, both {\it per se} and in
order to emphasize how crucial the contact approximation is in
this regime.

\begin{figure}
\begin{tabular}{cc}
  \includegraphics[width=180pt]{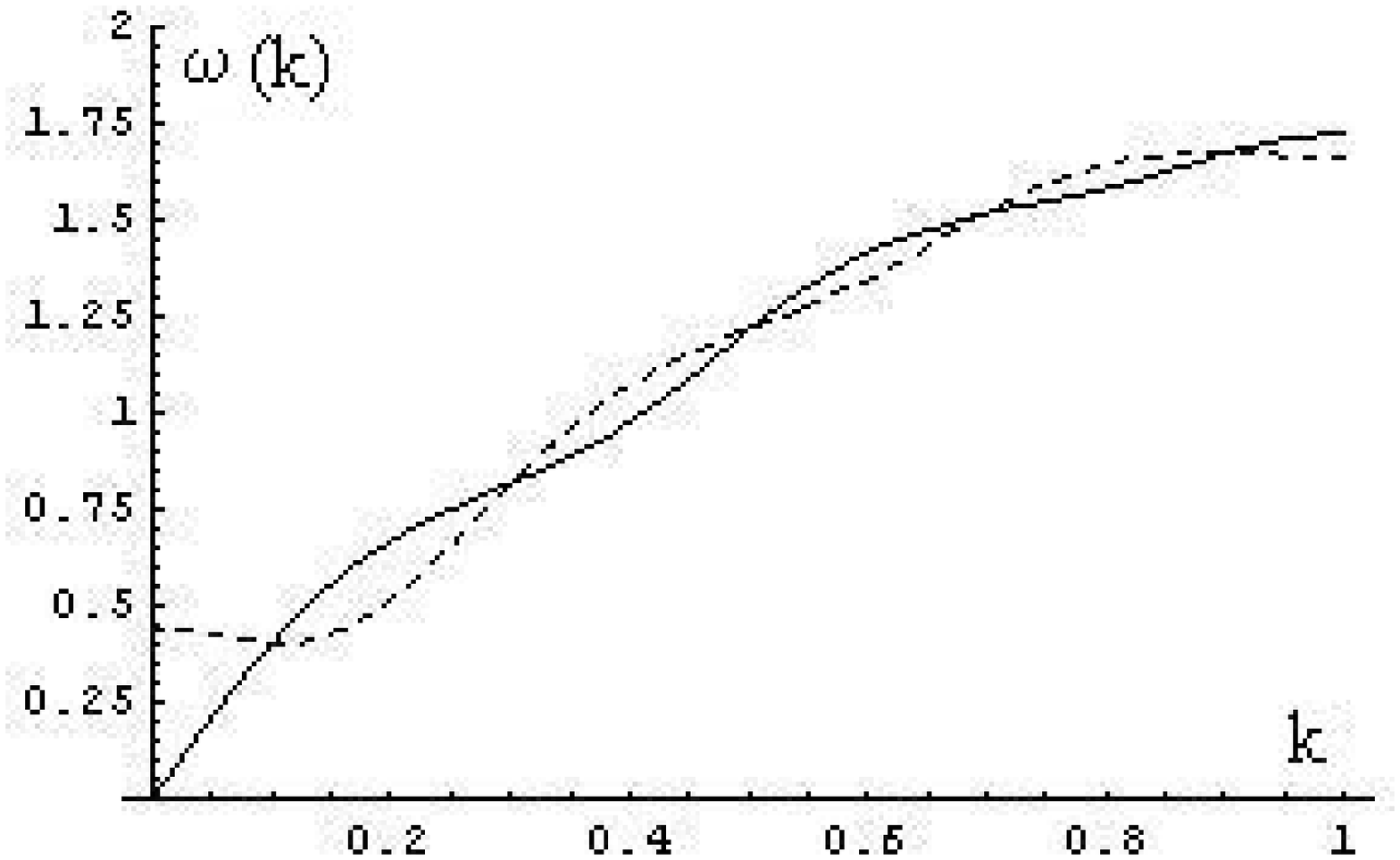} & (a) \\
  \includegraphics[width=180pt]{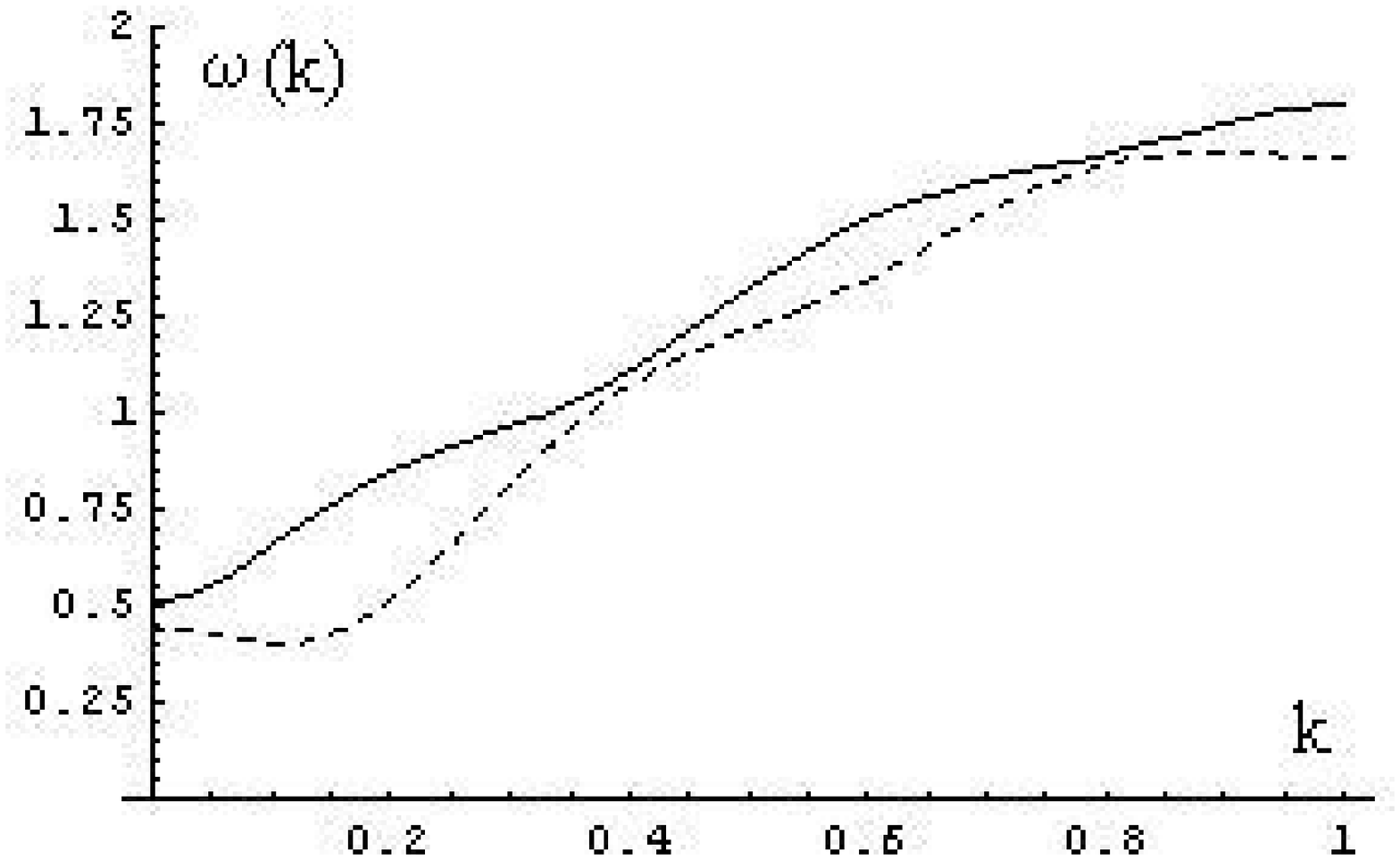} & (b) \\
  \includegraphics[width=180pt]{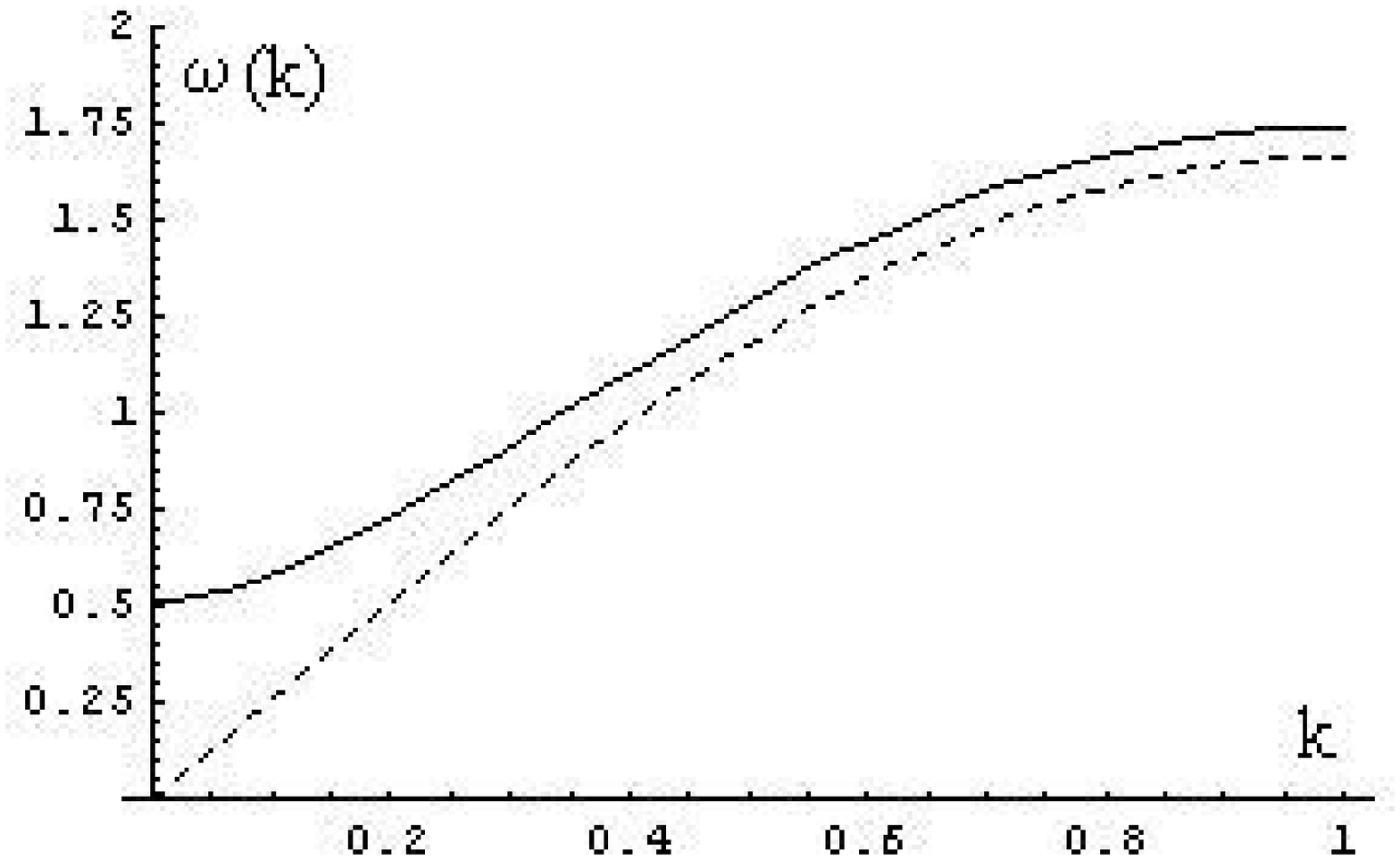} & (c) \end{tabular}
  \caption{Dispersion relations for different versions of the \Y
  model.
  (a): The dispersion relations for $\om = \om (k)$ as described by (A.3)
  for the helicoidal Y model without the contact approximation.
  (b): Dispersion relations for the standard (i.e. $\ell_0 = 0$) Y model
  with helicoidal terms. (c): Dispersion relations for the standard
  Y model without helicoidal terms.
  In all plots, the continuous line refers to the $\psi$ branch,
  the dashed one to the $\chi$ branch. We have used the values
  $K_s = 0.13 {\rm eV/rad}^2$, $K_h = 0.009 {\rm eV/rad}^2$
  \cite{GaeJBP}; here $k$ is given in units of $\pi / \de$, and $\om$ in ${\rm
  ps}^{-1}$. The $\chi$ branch of the standard Y model without helicoidal terms
  (plot on the right) coincides with
  the degenerate dispersion relations of the Y model without the contact
  approximation and without helicoidal terms.} \end{figure}

Small amplitude dynamics around the equilibrium position $\psi_n =
\chi_n = 0$ is described by the linearization of (1.10) at
$(0,0)$. As $V_0$, hence $V$, is non quadratic there, these
collapse to a pair of identical equations for $\psi$ and $\chi$.
In particular, the dispersion relations are now completely
degenerated, and the intrapair term has no role in them.

This degeneration is removed by introducing in the model the
``helicoidal'' terms mentioned in sect.1
\cite{Dauhel,Gaehel,GRPD}. This amounts to introducing in the
Lagrangian a new potential term
$$ U_h \ = \ {K_h \over 2} \ \sum_i \[ \( \vth^{(1)}_{i+p} -
\vth^{(2)}_{i}\)^2 \, + \, \( \vth^{(2)}_{i+p} -
\vth^{(1)}_{i}\)^2 \] \ . \eqno(A.1) $$ Here $p$ is the half-pitch
of the helix in nucleotide units; it takes the value $p=5$ in
B-DNA.

The introduction of this terms in $L$ entails that a new term
should be added to the right hand side of the Euler-Lagrange
equations (1.10); this new term is linear and thus is also present
in their linearization.

With standard computations, the new linearized equations are (note
the sign differences in the new terms)
$$ \begin{array}{rl}
I {\ddot \psi}_n \ =& \ K_s \, \( \psi_{n+1} - 2 \psi_n +
\psi_{n-1} \) \ + \\ & + \ K_h \, \( \psi_{n+p} - 2 \psi_n + \psi_{n-p} \) \ , \\
I {\ddot \chi}_n \ =& \ K_s \, \( \chi_{n+1} - 2 \chi_n +
\chi_{n-1} \) \ + \\ & - \ K_h \, \( \chi_{n+p} + 2 \chi_n +
\chi_{n-p} \) \ . \end{array} \eqno(A.2) $$

Passing to the continuum approximation and Fourier transforming
via $ \Phi (x,t) = f_{k \om} \exp[i (k x + \om t)]$, $\Xi (x,t) =
g_{k \om} \exp[i (k x + \om t)]$, the above yield
$$ \begin{array}{rl}
\om^2_\psi  =&  (2 K_s/I)  [ 1 - \cos (k \de ) ]  +  (2 K_h
/I)  [ 1 - \cos (k p \de ) ] \\
\om^2_\chi  =&  (2 K_s /I)  [ 1 - \cos (k \de ) ] + (2 K_h /I) [ 1
+ \cos (k p \de ) ] \end{array} \eqno(A.3) $$ for the $\psi$ and
the $\chi$ branch respectively (note that here the continuous wave
number $k$ has the dimension of $[L]^{-1}$, and the lattice
spacing $\de$ sets the space scale; one could of course also pass
to a dimensionless wave number $\kappa = k \de$). These are the
dispersion relations -- plotted in fig.5a -- for the helicoidal \Y
model without the contact approximation. Note that a nonzero
$\ell_0$ causes the presence, even in the helicoidal case, of
phonon modes (contrary to the $\ell_0 = 0$ case).

For comparison purposes, we note that the dispersion relations for
the standard \Y model are
$$ \begin{array}{rl}
\om^2_\psi =& (2 K_s/I)  [ 1 - \cos (k \de ) ] +  (2 K_h
/I)  [ 1 - \cos (k p \de ) ] + \\ &  + 2 K_p r^2 /I \\
\om^2_\chi =& (2 K_s /I)  [ 1 - \cos (k \de ) ] + (2 K_h /I)  [ 1
+ \cos (k p \de ) ] \end{array} \eqno(A.4) $$ (the non-helicoidal
case is obtained setting $K_h$ in the above); these are plotted in
fig.5b, and in fig.5c for the non-helicoidal ($K_h=0$) case.

Quantities of physical interest are readily evaluated, at least
numerically, from (A.3). For $K_h \not=0$, the $\chi$ branch has a
nontrivial minimum; this is reached for $k = k_0 = 0.101
\AA^{-1}$, i.e. for $\la = \la_0 = 62 \AA = 18 \de$. It is
suggestive to remark that if solitons grow out of oscillations
triggered by thermal excitations, and assuming equipartition of
energy between Fourier modes, the larger amplitude excitations
have a characteristic size which is of the order of the size of
the transcription bubble, the latter being $15-20$ base pairs
\cite{GRPD,YakuBook}. To the above value of $k_0$ corresponds a
frequency $\om_0 = 0.4 {\rm ps}^{-1}$, i.e. a period of
oscillations $T_0 \simeq 16 {\rm ps}$; typical values for $T_0$
observed in experiments are in the picoseconds range.

The speed of phonon excitations $v_s$ corresponds to $d \om_\psi
(k) / d k$ for $k \to 0$. In the $\ell_0 = 0$ case, with the
values of parameters given above, we get $v_s \simeq 280$m/s,
while if we drop the contact approximation we get $v_s \simeq
470$m/s. These values should be compared with experimental
observations for the speed of torsional waves, which provide
values in the range of $v_s = 1.6 \pm 0.3 $ Km/s \cite{YakuBook}.


\begin{thebibliography}{39}

\bibitem{BCP} M. Barbi, S. Cocco and M. Peyrard, ``Helicoidal model for DNA
opening'', {\it Phys. Lett. A} {\bf 253} (1999), 358-369

\bibitem{BCPR} M. Barbi, S. Cocco, M. Peyrard and S. Ruffo, ``A twist-opening
model of DNA'', {\it J. Biol. Phys.} {\bf 24} (1999), 97-114

\bibitem{BLPT} M. Barbi, S. Lepri, M. Peyrard and N.
Theodorakopoulos, ``Thermal denaturation of a helicoidal DNA
model'', {\it Phys. Rev. E} {\bf 68} (2003), 061909

\bibitem{BCG} R.K. Bullough, P.J. Caudrey and H.M. Gibbs, ``The double sine-Gordon
equations'', pp. 107-141 in {\it Solitons} (R.K. Bullough and P.J.
Caudrey eds.), Springer (Berlin) 1980

\bibitem{CDG} M. Cadoni, R. De Leo and G. Gaeta, ``A composite Y model of DNA
dynamics'', preprint {\it q-bio.BM/0604014} (2006)

\bibitem{CD} C. Calladine and H. Drew, {\it Understanding DNA},
Academic Press (London) 1992; C. Calladine, H. Drew, B. Luisi and
A. Travers, {\it Understanding DNA} ($3^{rd}$ edition), Academic
Press (London) 2004

\bibitem{CDeg} F. Calogero and A. Degasperis, {\it Spectral Transform and Solitons:
tools to solve and investigate nonlinear evolution equations},
North-Holland (New York) 1982

\bibitem{Dauhel} Th. Dauxois, ``Dynamics of breather modes in a
nonlinear helicoidal model of DNA'', {\it Phys. Lett. A} {\bf 159}
(1991), 390-395

\bibitem{Eng} S.W. Englander, N.R. Kallenbach, A.J. Heeger, J.A.
Krumhansl and A. Litwin, ``Nature of the open state in long
polynucleotide double helices: possibility of soliton
excitations'', {\it PNAS USA} {\bf 77} (1980), 7222-7226

\bibitem{Gaehel} G. Gaeta, ``On a model of DNA torsion dynamics'',
{\it Phys. Lett. A} {\bf 143} (1990), 227-232

\bibitem{GaeJBP} G. Gaeta, ``Results and limits of the soliton
theory of DNA transcription'', {\it J. Biol. Phys.} {\bf 24}
(1999), 81-96

\bibitem{GRPD} G. Gaeta, C. Reiss, M. Peyrard and Th. Dauxois,
``Simple models of non-linear DNA dynamics'', {\it Rivista del
Nuovo Cimento} {\bf 17} (1994) n.4, 1--48

\bibitem{GML} J.A. Gonzalez and M. Martin-Landrove, ``Solitons in
a nonlinear DNA model'', {\it Phys. Lett. A} {\bf 191} (1994),
409-415

\bibitem{KMP} Y. Kafri, D. Mukamel and L. Peliti, ``Why is the DNA
denaturation transition first order?'', {\it Phys. Rev. Lett.}
{\bf 85} (2000), 4988-4991; ``Melting and unzipping of DNA'', {\it
Eur. Phys. J. B} {\bf 27} (2002), 135-146

\bibitem{PeyNLN} M. Peyrard, ``Nonlinear dynamics and statistical
physics of DNA'', {\it Nonlinearity} {\bf 17} (2004) R1-R40

\bibitem{PB} M. Peyrard and A.R. Bishop, ``Statistical mechanics
of a nonlinear model for DNA denaturation'', {\it Phys. Rev.
Lett.} {\bf 62} (1989), 2755-2758

\bibitem{Vol} M. Volkenstein, {\it Biophysique}, MIR (Moscow) 1985

\bibitem{YakPLA} L.V. Yakushevich, ``Nonlinear DNA dynamics: a new
model'', {\it Phys. Lett. A} {\bf 136} (1989), 413-417

\bibitem{YakPhD} L.V. Yakushevich, ``Nonlinear DNA dynamics:
hyerarchy of the models'', {\it Physica D} {\bf 79} (1994), 77-86

\bibitem{YakuBook} L.V. Yakushevich, {\it Nonlinear Physics of
DNA}, Wiley (Chichester) 1998; second edition 2004

\bibitem{YakPRE} L.V. Yakushevich, A.V. Savin and L.I. Manevitch, ``Nonlinear dynamics
of topological solitons in DNA'', {\it Phys. Rev. E} {\bf 66}
(2002), 016614

\bibitem{ZC} F. Zhang and M.A. Collins, ``Model simulations of DNA
dynamics'', {\it Phys. Rev. E} {\bf 52} (1995), 4217-4224

\end{thebibliography}
\end{document}